\documentclass[aps,pra,floatfix,twocolumn,superscriptaddress,tightenlines]{revtex4}

\newcommand{\ket}[1]{\left\vert#1\right\rangle}

\newcommand{\pr}{\partial}

\usepackage{amsmath}
\usepackage{graphicx}
\usepackage{psfrag}
\usepackage[usenames]{color}
\usepackage{hyperref}
\usepackage{bm}
\usepackage{soul}
\usepackage{rotating}
\newcommand{\unit}[1]{\ensuremath{\;\mathrm{#1}}}
\usepackage{lscape}
\usepackage{float}

\begin{document}

\title{Two-photon quantum walks in an elliptical direct-write waveguide array}

\author{J.~O.~Owens}
\affiliation{ARC Centre for Engineered Quantum Systems, ARC Centre for Quantum Computation and Communication Technology, School of Mathematics and Physics, University of Queensland, 4072 Brisbane, QLD, Australia}

\author{M.~A.~Broome}
\affiliation{ARC Centre for Engineered Quantum Systems, ARC Centre for Quantum Computation and Communication Technology, School of Mathematics and Physics, University of Queensland, 4072 Brisbane, QLD, Australia}

\author{D.~N.~Biggerstaff}
\affiliation{ARC Centre for Engineered Quantum Systems, ARC Centre for Quantum Computation and Communication Technology, School of Mathematics and Physics, University of Queensland, 4072 Brisbane, QLD, Australia}

\author{M.~E.~Goggin}
\affiliation{Department of Physics, Truman State University, Kirksville, Missouri 63501, USA}

\author{A.~Fedrizzi}
\affiliation{ARC Centre for Engineered Quantum Systems, ARC Centre for Quantum Computation and Communication Technology, School of Mathematics and Physics, University of Queensland, 4072 Brisbane, QLD, Australia}

\author{T.~Linjordet}
\affiliation{ARC Centre for Engineered Quantum Systems, Department of Physics and Astronomy, Macquarie University, North Ryde NSW 2109, Australia}

\author{M.~Ams}
\affiliation{ARC Centre for Ultrahigh bandwidth Devices for Optical Systems, Centre for Quantum Science and Technology, MQ Photonics Research Centre, Department of Physics and Astronomy, Macquarie University, North Ryde NSW 2109, Australia}

\author{G.~D.~Marshall}
\affiliation{ARC Centre for Ultrahigh bandwidth Devices for Optical Systems, Centre for Quantum Science and Technology, MQ Photonics Research Centre, Department of Physics and Astronomy, Macquarie University, North Ryde NSW 2109, Australia}

\author{J.~Twamley}
\affiliation{ARC Centre for Engineered Quantum Systems, Department of Physics and Astronomy, Macquarie University, North Ryde NSW 2109, Australia}

\author{M.~J.~Withford}
\affiliation{ARC Centre for Ultrahigh bandwidth Devices for Optical Systems, Centre for Quantum Science and Technology, MQ Photonics Research Centre, Department of Physics and Astronomy, Macquarie University, North Ryde NSW 2109, Australia}

\author{A.~G.~White}
\affiliation{ARC Centre for Engineered Quantum Systems, ARC Centre for Quantum Computation and Communication Technology, School of Mathematics and Physics, University of Queensland, 4072 Brisbane, QLD, Australia}

\date{\today}
\begin{abstract}
Integrated optics provides an ideal test bed for the emulation of quantum systems via continuous-time quantum walks. Here we study the evolution of two-photon states in an elliptic array of waveguides. We characterise the photonic chip via coherent-light tomography and use the results to predict distinct differences between temporally indistinguishable and distinguishable two-photon inputs which we then compare with experimental observations. Our work highlights the feasibility for emulation of coherent quantum phenomena in three-dimensional waveguide structures.
\end{abstract}

\maketitle

\section{Introduction}
Computer simulations of complex systems have had spectacular success in modern science due to sophisticated classical approximation methods and steadily increasing computational power. However, it is ultimately impractical to accurately simulate the dynamics of even moderately-sized quantum systems due to their exponentially-increasing parameter space. Quantum simulation, as first proposed by Feynman \cite{feynman1982spc}, offers a path forward whereby the simulator itself operates via quantum instead of classical dynamics.

There are two possible approaches to modeling a quantum system. The first is \emph{simulation}, in which a digital model of a system's Hamiltonian yields a physical quantity. An example is provided by recent quantum chemistry simulations of a Hydrogen molecule~\cite{lanyon2010tqc,du2010nim} on a small-scale quantum computer. The second is \emph{emulation}, which involves the modeling of a quantum system with a better-controllable system which is governed by a sufficiently similar Hamiltonian~\cite{lloyd1996uqs}.

Quantum walks---natural extensions of the ubiquitous classical random walk into the quantum world~\cite{aharonov1993qrw,kempe2003qrw}---provide an ideal framework for emulation due to their rich dynamics~\cite{mulken2011, plenio2008, mohseni2008eqw}. There are two limiting cases, discrete and continuous. In discrete-time quantum walks, one or more (interacting) quantum particles (the so-called quantum coin) evolve on a graph, with their evolution governed by their internal quantum (coin) states. The discrete-time quantum walk on a line is the best studied example of such a walk and it has been demonstrated in a number of physical systems \cite{karski2009qwp,zaehringer2010rqw,schmitz2009qwt,schreiber2010pwl,broome2010dsp}. 

In continuous-time quantum walks, in contrast, there are no coin operations, the evolution is defined entirely in position space \cite{farhi1998qcd}. These walks require a well-controlled, continuous coupling between vertices, or lattice sites. Ideally, this coupling should be stable and robust against decoherence. Integrated optics is perfectly suited for this task. Lithographically written, evanescently coupled surface waveguides have already been used to study continuous time quantum walks on a line \cite{bromberg2009qcc}. A disadvantage of surface waveguides is that they can only realise simple, one-dimensional graphs with limited interconnectivity. 

Physically more interesting three-dimensional structures can be engineered in laser-written optical waveguide arrays in dielectric materials such as fused silica~\cite{marshall2009lww}. The two-dimensional graphs that can be realised with this technique allow the study of new quantum walk phenomena, such as wave communication \cite{hein2009wca}, cooperative quantum games \cite{bulger2008pdc} and the creation of topological phases in two dimensions \cite{kitagawa2010etp}. Examples of waveguide array structures which have been manufactured with this technique include rings, regular and hexagonal lattices, X-shapes and triangular shapes \cite{szameit2007cwa,roepke2007tdh,keil2010pct}. To date, however, these have only been explored with coherent light. True quantum effects, and associated genuine improvements in efficiency of emulating quantum systems will only emerge for non-classical input states \cite{rohde2011mdt}, such as recently demonstrated experimentally for a one-dimensional waveguide lattice in \cite{peruzzo2010qwc}.

Here, we study multi-walker continuous-time quantum walks in an optical chip containing an elliptical arrangement of coupled direct-write waveguides. We characterise the optical chip via coherent light tomography, effectively realising single particle walks 
and use the results to predict correlations in non-classical two-photon walks which we compare with experimental observations. Our work is an important step towards the emulation of quantum systems in three-dimensional integrated photonic architectures.

\begin{figure}[t!]
\centering
\includegraphics[width=1\columnwidth]{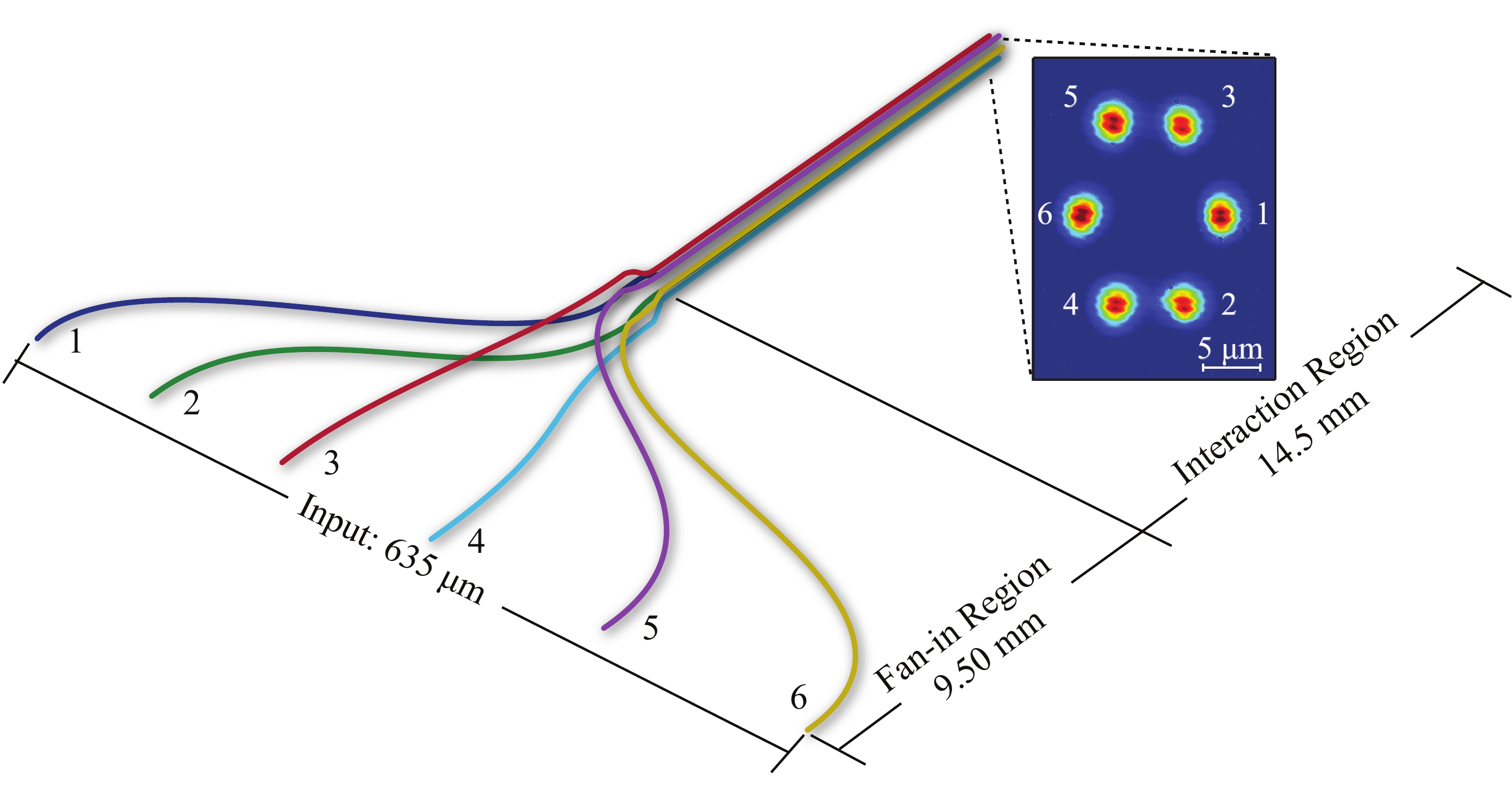}
\caption{Schematic of the integrated waveguide circuit and associated output pattern. The fan-in region here is scaled by a factor of $\sim54$ relative to the interaction region along the beam direction. At the input the waveguides are equally spaced at 127\unit{\mu m}. They then converge via a two-stage fan-in to their eventual elliptical configuration, which is shown in the CCD camera picture in the inset: the waveguides are arranged with equal angular spacing around an ellipse with semi-major and semi-minor radii of 10.2 and 7.0~$\mu$m respectively.}
\label{fig:circuit}
\end{figure}

\section{Device Description}

The circuit for the continuous time quantum walk, shown in figure~\ref{fig:circuit}, consists of six waveguides written into a chip of high-purity fused silica using an ultrafast direct-write technique~\cite{marshall2009lww,davis1996, nolte2003}. In this technique, femtosecond Ti:Sapphire laser pulses tightly focused inside the sample yield localized refractive index modifications.
By translating the sample in three dimensions these index modifications can manifest as waveguides tracing three-dimensional curves, a process that cannot be replicated in conventional lithographic waveguide fabrication techniques. Our chip was written with a 1~kHz repetition rate, 800~nm, 120~fs laser, passed through a 520~$\mu$m slit and focused with a 40x 0.6 NA microscope objective.

At the circuit input, the six waveguides are arranged along a line with an equal spacing of 127
$\mu$m. This allows each waveguide to be addressed individually and simultaneously. The waveguides then converge via a two-stage fan-in to their final configuration as shown in figure~\ref{fig:circuit}. In the primary fan-in stage, which occupies the first 8.5~mm of the chip, the waveguides follow S-bend curves from their linear input arrangement to an elliptical configuration twice as large in radius as their final configuration. In the second fan-in stage, comprising the next 1~mm, further S-bends shrink this array to its final geometry of an ellipse with a semi-major axis of 10.2~$\mu$m and semi-minor axis of 7.0~$\mu$m. All the S-bends are of the `raised-sine' form which has been shown to minimise bend loss \cite{minford1982}, while the two-stage fan-in configuration was designed to minimise coupling (in particular, \emph{asymmetrical} coupling) between non-adjacent waveguides before they reach their final configuration.

The light evolution in such a waveguide array is governed by the evanescent inter-waveguide coupling which drops off exponentially as a function of the waveguide distance. It is theoretically described by a coupled-oscillator Hamiltonian, which is described in the Appendix.

\begin{figure}[b]
\centering
\includegraphics[width=1\columnwidth]{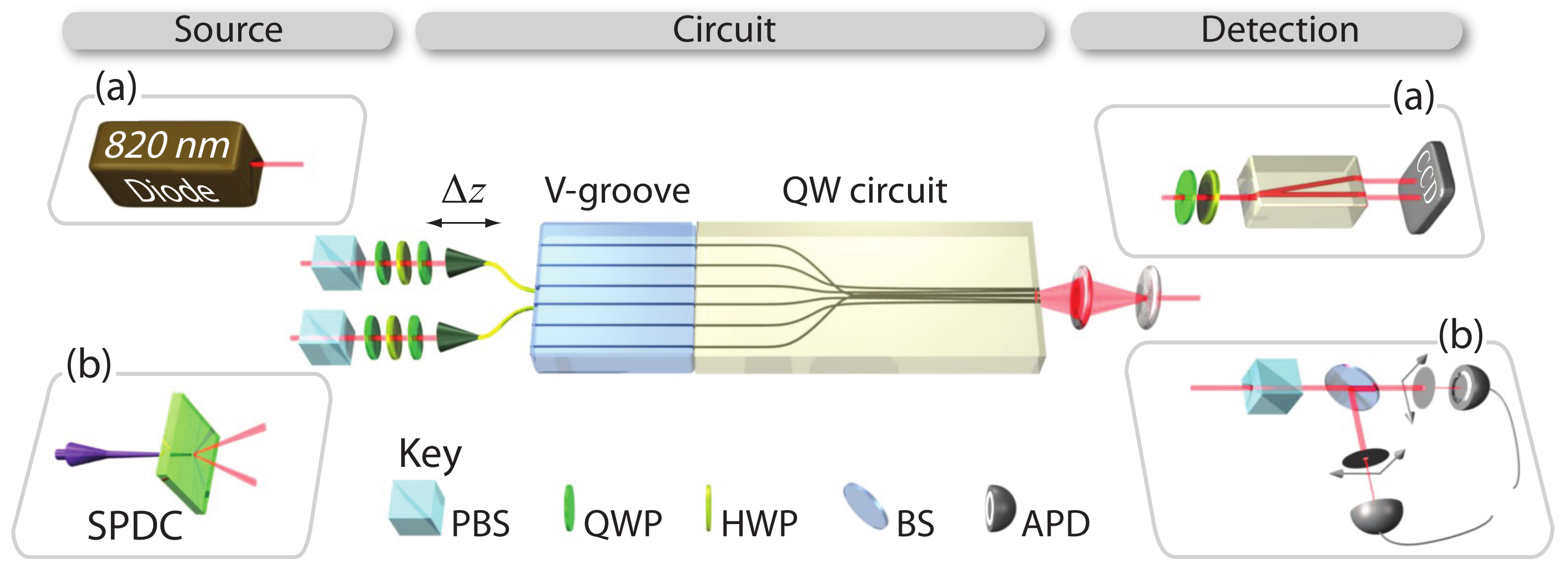}
\caption{Experimental setup. The chip can be addressed with 6 individual single mode fibres. The input light polarisation is set by a combination of half- and quarter-wave plates (HWP, QWP) and polarising beamsplitters (PBS). The chip output is magnified and then collimated with two spherical lenses. (a) The chip was characterised with an $820$~nm laser diode, imaged onto a CCD camera via a polarising prism. (b) Quantum walks were performed with two-photon inputs created via spontaneous parametric downconversion (SPDC). The relative delay $\Delta z$ between the two input photons was adjusted using a translation stage. We used a 50/50 beamsplitter (BS) and two $\sim  \! 500~\mu$m apertures to select a combination of output ports and measure the two-photon correlations in coincidence using single-photon avalanche diodes.} \label{fig:tpwalk}
\end{figure}

\section{Optical chip characterisation}

\begin{figure}[t]
\centering
\includegraphics[width=1\columnwidth]{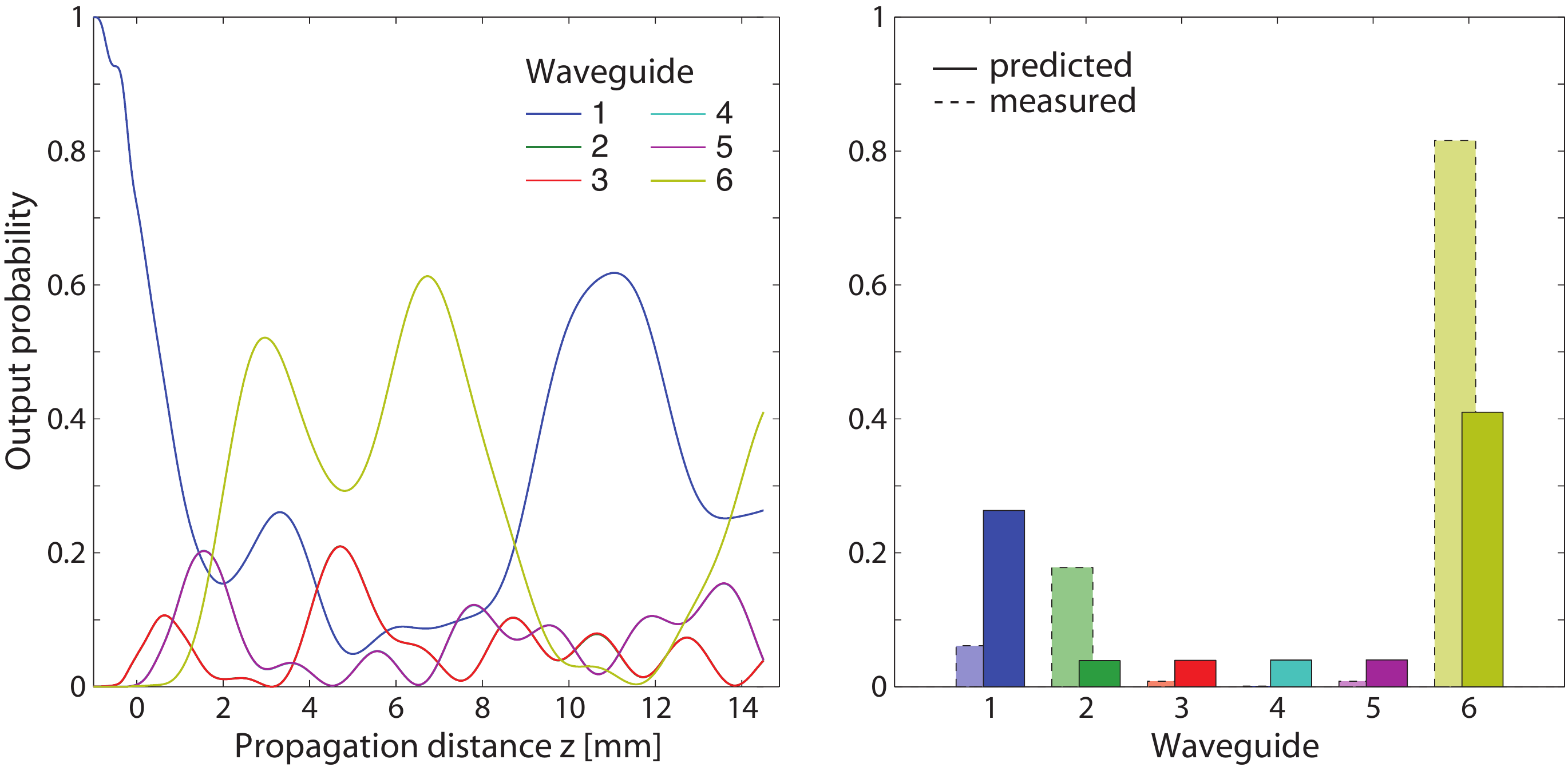}
\caption{Comparison of numerical simulations and observed probability distributions for the optical chip in figure~\ref{fig:circuit}. (a) The propagation dynamics predicted using optical waveguide simulation software (see Appendix) with light input into waveguide $1$, as a function of $z$. The simulation includes modelling of inter-waveguide coupling during the $1$~mm second fan-in stage at $z{\leq}0$. The curves for waveguides 2 and 4, and 3 and 5 overlap due to symmetry. (b) The predicted distribution at the output of the circuit, and the corresponding observed probabilities. The distinct differences in the two distributions indicate that the optical response of the chip is not scalar, instead suggesting significant vectorial behaviour. This was confirmed by further tomographic analysis.}
\label{fig:propdists}
\end{figure}

The experimental setup is shown in figure~\ref{fig:tpwalk}. Light was coupled into the chip via a V-groove array, which houses six single-mode input fibres on a line, matching the input spacing of the circuit waveguides. We first illuminated individual waveguides in the chip using coherent light from an $820$~nm laser diode, see figure~\ref{fig:tpwalk}a). The output intensity profiles were processed in Matlab and converted into probability distributions. When compared to the numerical simulation obtained with the software suite which was used to design the chip, the predicted distribution at
the circuit output was starkly different to the observed distribution as shown in figure~\ref{fig:propdists}. 
This behaviour prompted us to empirically determine the full optical response of the circuit using six-port, polarisation-sensitive coherent light tomography.

For this tomography the input polarisation was set using a bare reference fibre on top of the chip. Into each waveguide we input the following set of polarisation states: $\{\ket{H},\ket{V},\ket{D},\ket{A},\ket{L},\ket{R}\}$, where $\ket{H}$ and $\ket{V}$ represent horizontal and vertical polarisation and $\ket{D/A}{=}(\ket{H}\pm\ket{V})/{\sqrt{2}}$, $\ket{L/R}{=}(\ket{H}\pm i\ket{V})/{\sqrt{2}}$. For each output we measured the same six polarisation components, obtaining a total of $216$ CCD images.

Using output intensity distributions taken from these images, we subsequently reconstructed an array of 36 Mueller matrices $\mathcal{M}$~\cite{hecht1974o}. This array completely characterises the optical response of the circuit, quantifying the effects of three distinct processes: notably polarisation-dependent inter-waveguide coupling, birefringence, and polarisation-dependent loss~\cite{hecht1974o}.

\begin{figure}[b!]
\centering
\includegraphics[width=1\columnwidth]{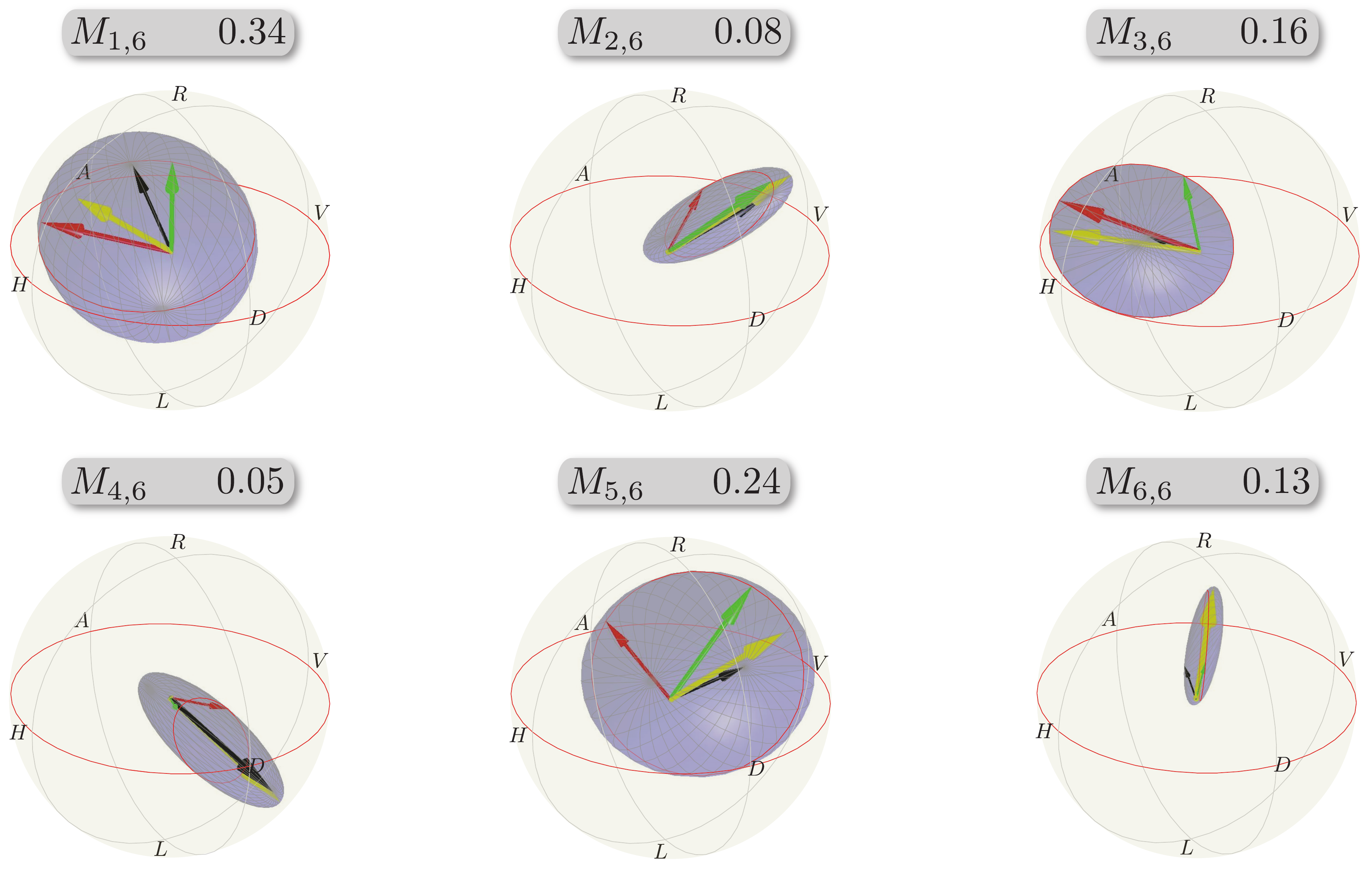}
\caption{Poincar{\'e} sphere representation of the corresponding Mueller matrices, $M_{out,in}$ (see Appendix) describing the transformation from input waveguide $6$ to outputs $1-6$. Input states shown on the outer Poincar{\'e} spheres are mapped to different locations  on the blue ellipsoids, simultaneously showing polarisation-dependent coupling and birefringent effects in the respective channel. The relative orientation of the ellipsoids is indicated by the set of orthogonal states $\ket{H}$ (red arrow), $\ket{D}$ (green arrow) and $\ket{R}$ (black arrow), and the point of contact between sphere and ellipsoid is indicated by the yellow arrow. Note that the arrow lengths are proportional to output power, \emph{not} degree of polarisation. The numbers above each sphere give the normalised average power coupled into the respective channel.}
\label{fig:ellipsoids}
\end{figure}

The results indicate strong birefringence in each of the six waveguides. Most notably, when the state $\ket{H}$ is input into waveguide $5$, $29$\% of the overall output state across the six channels is rotated to $\ket{V}$. For the input channels $1$, $2$ and $4$, the overall polarisation rotation was small, with more than $91\%$ of $\ket{H}$ being retained in each case. Furthermore, there was significant polarisation-dependent coupling between the waveguides for all input channels. For instance, for input $\ket{H}$ into waveguide $1$, $80$\% of the total output intensity was observed in channel $6$, however when the input state was $\ket{V}$, only $11$\% of the total output intensity was contained in this channel. An exemplary selection of Mueller matrices,  illustrated on Poincar{\'e} spheres and quantifying these effects is shown in figure~\ref{fig:ellipsoids}. The full matrix array $\mathcal{M}$ can be found in the Appendix.

In addition, the whole chip exhibited significant polarisation dependent loss; integrating over all output channels, we observed an excess $38\%$ loss of $\ket{V}$ compared to $\ket{H}$ for waveguide $6$. This may be due to a combination of absorption into the bulk of the circuit, or polarisation dependence of the input coupling efficiency from the V-groove array to the chip, or both.

\section{Two-photon walks}
The coherent-light tomography encompasses all possible single-photon walks in this chip, since a single photon shares the coherence properties of a coherent light beam. However, these walks can be efficiently simulated classically and it is not until we input multi-photon states that we observe quantum effects \cite{rohde2011mdt}. 

A schematic of the setup for the two-photon walk experiment is shown in figure~\ref{fig:tpwalk}b). Pairs of single photons are created via spontaneous parametric down-conversion: a mode-locked $76$~MHz laser centred at $820$~nm is frequency doubled to $410$~nm and focused into a $2$~mm long $\beta$-barium borate (BBO) crystal, phase-matched for type-I downconversion. After passing through interference filters at $820\pm2.5$~nm, the degenerate output photon pairs are coupled into single-mode fibres equipped with manual polarisation controllers. They are then connected to the V-groove fibre array and thence to the chip.

We created photon pairs at a measured rate of $180$~kHz, of which we detected a total of $6.5$~kHz at the circuit output. The main loss contributions stemmed from the poor input coupling efficiency between the V-groove array and the chip $({\sim}31\%)$, and the 50/50 beamsplitter $(\sim50\%)$ at the output. The imperfect coupling is mostly due to a slight mismatch between the spacing of the fibres in the V-groove array and the locations of the input ports of the circuit. The observed loss could be significantly reduced by using a more sophisticated imaging system.

\begin{figure}[t!]
\centering
\includegraphics[width=1\columnwidth]{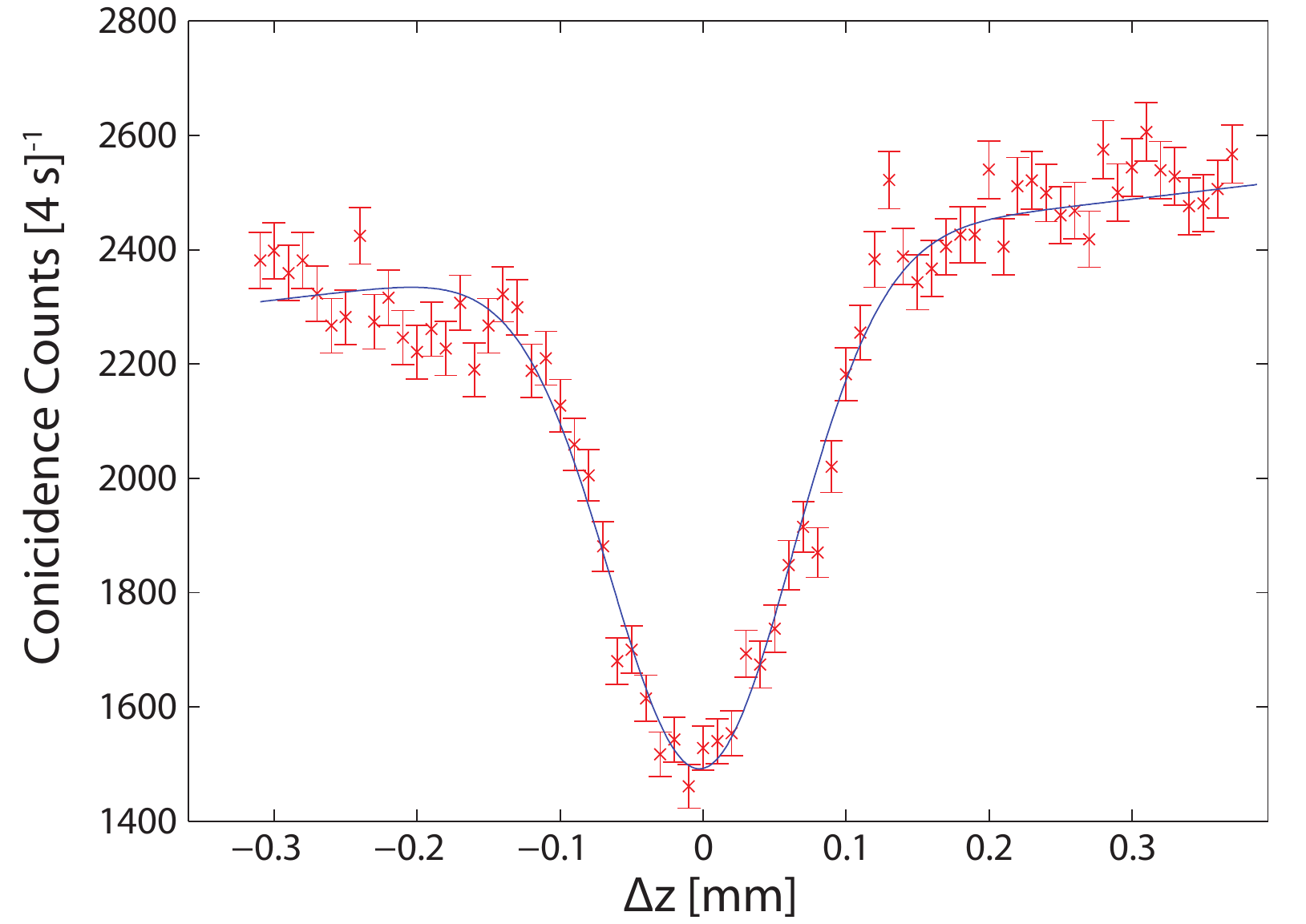}
\caption{Example of observed two-photon interference between output waveguides 2 and 6 as a function of relative path difference between photon pairs input into neighbouring waveguides $1$ and $2$. The visibility of the dip is $\mathcal{V}_{2,6} =38 \pm 2\%$, calculated from a Gaussian fit (blue line).}
\label{fig:hom12in26out}
\end{figure}

\begin{figure}[b!]
\centering
\includegraphics[width=1\columnwidth]{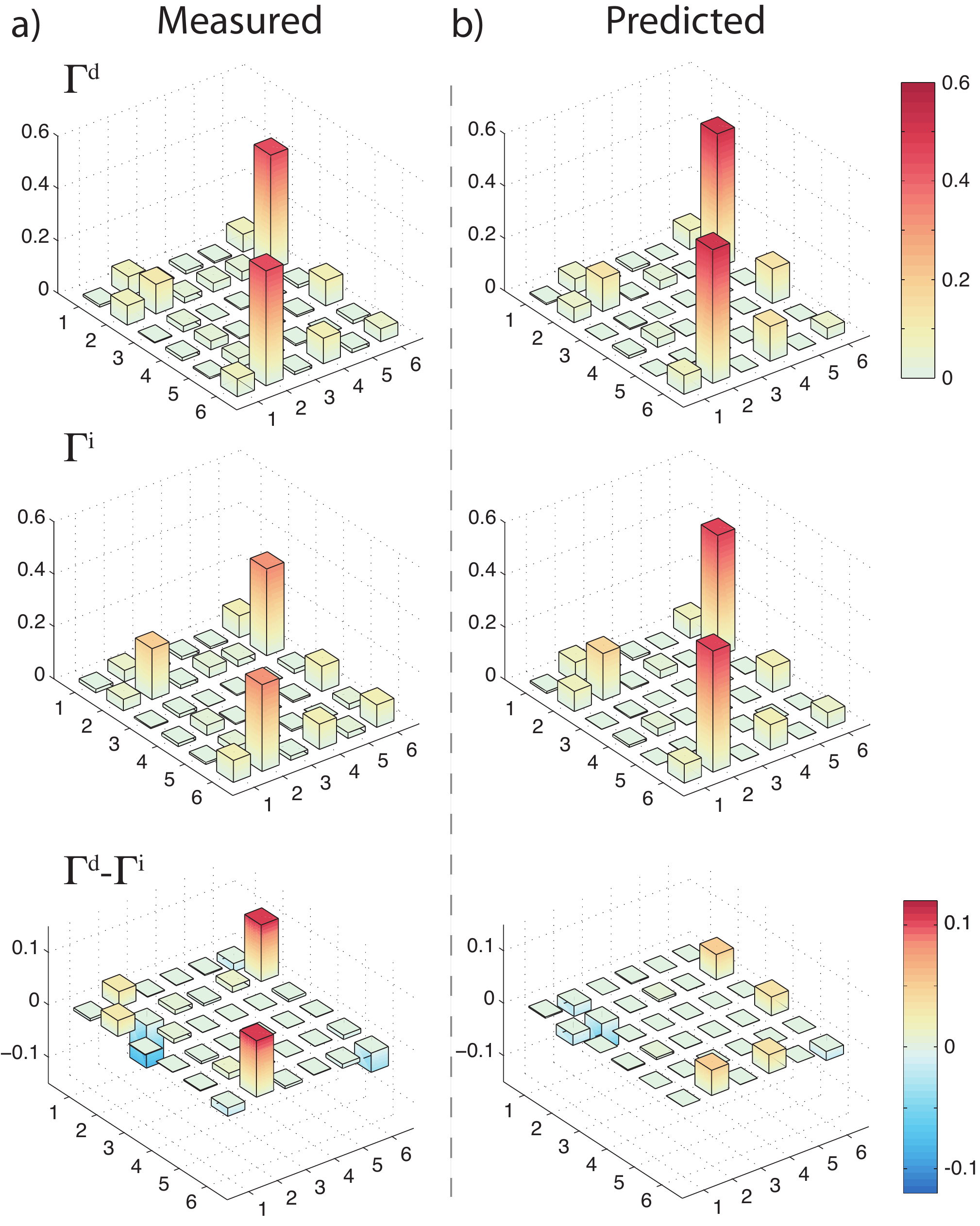}
\caption{Correlation matrices for nearest-neighbour input channels $1$ and $2$. We recorded the photon-coincidence counts at each of the $36$ pairs of output channels in a $20$ second time-window. (a) The measured and (b) predicted correlation matrices for (left) temporally distinguishable photon pairs $\Gamma^{d}$, (center) temporally indistinguishable simultaneous walkers $\Gamma^{i}$, and (right) the difference $\Gamma^{d} - \Gamma^{i}$. The coincidence probability at the outputs $2$ and $6$ between the two plots reflects the two-photon interference dip shown in figure \ref{fig:hom12in26out}. The measured uncertainties are not plotted, as they are too small to be seen on the plots.}
\label{fig:corr12total}
\end{figure}

We carried out two-photon quantum walks in two separate scenarios: with temporally distinguishable and indistinguishable photons. When the photons entering the chip are temporally distinguishable, i.e. with a time delay larger than their respective coherence times, they perform independent quantum walks with local evolution. When they enter the chip simultaneously, $\Delta z=0$, they experience non-classical two-photon interference~\cite{hong1987hom}, which can yield quantum dynamics, including generation of two-photon entanglement. The theoretic qualitative difference between these two cases is described in the Appendix. Figure~\ref{fig:hom12in26out} shows an exemplary calibration scan of coincidence counts as a function of time delay. The signature of indistinguishable quantum walkers manifests as a dip in the rate of coincident detection events, $C$, at zero delay, with an interference visibility of $\mathcal{V} = (C_{\max}-C_{\min})/C_{\max}$ of $38\pm2\%$. 

\begin{figure}[!h]
\centering
\includegraphics[width=1\columnwidth]{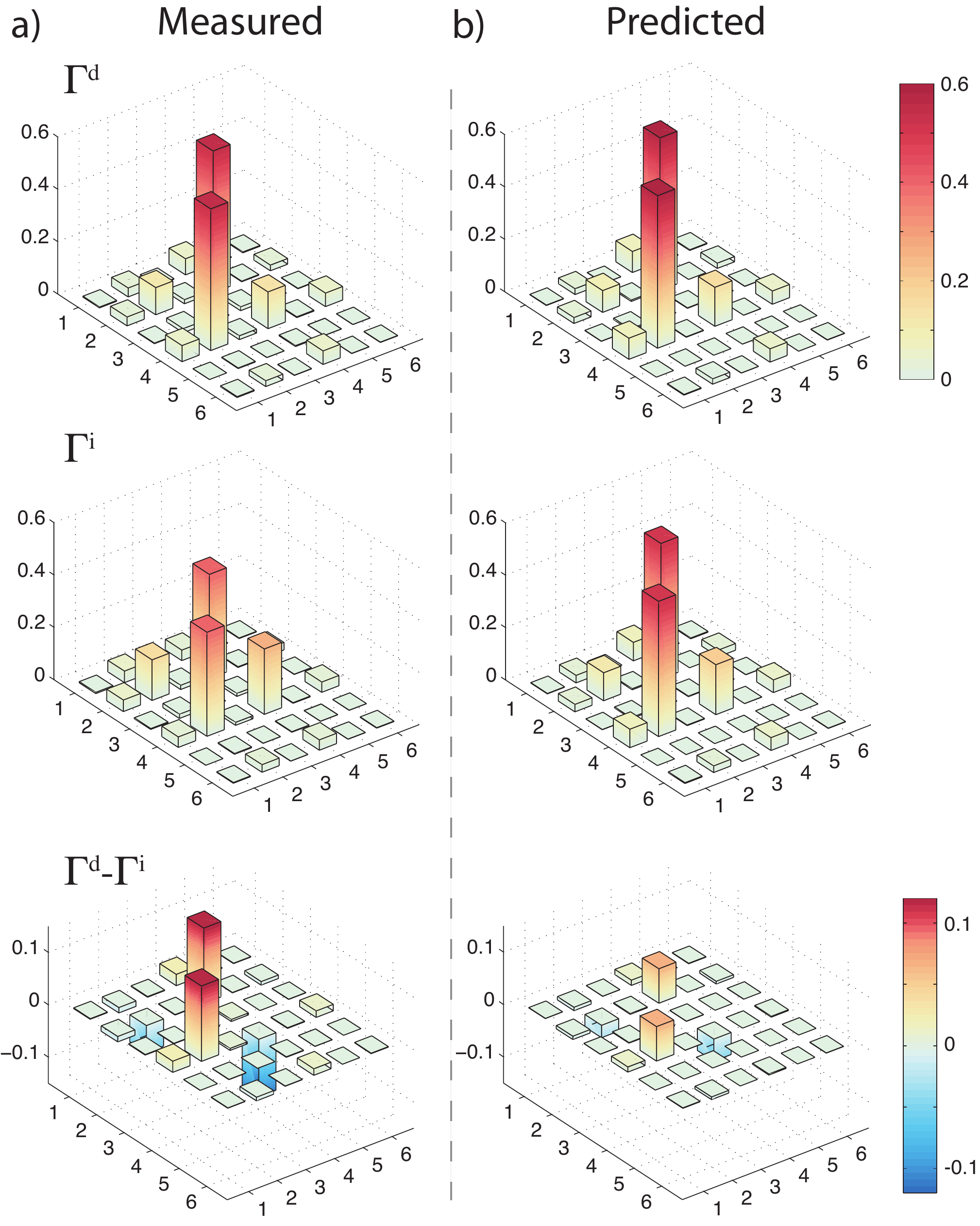}
\caption{Correlation matrices for next-nearest neighbour input channels $2$ and $4$. The plots follow the same convention as those in figure~\ref{fig:corr12total}, for both (a) the measured and (b) predicted correlation matrices. Uncertainties are again too small to be seen on the graphs.}
\label{fig:corr24total}
\end{figure} 
The results for two-photon quantum walks for distinguishable and indistinguishable photons input into the \emph{nearest}-neighbour channels $1$ and $2$ are shown in figure~\ref{fig:corr12total}a), as the normalised coincidence probability distributions, $\Gamma^{d}$ and $\Gamma^{i}$ respectively. Distinct differences are observed between the two cases, as suggested by the strong two-photon interference signature in figure~\ref{fig:hom12in26out}.

The measured distributions are compared with predictions, figure~\ref{fig:corr12total}b), which are based on determining the components of the waveguide array unitary $U$, for a particular input polarisation, see Appendix for details. The generalised overlap fidelities $S$, defined in the Appendix, between our measurements and predictions are $S^{d}{=}93.4 \pm 0.3\%$ for the distinguishable walkers, and $S^{i}{=}91.6\pm 0.4\%$ for the indistinguishable walkers.

Figure~\ref{fig:corr24total}a) shows correlation matrices for inputs $2$ and $4$ as an example of two-photon walks with \emph{next}-nearest neighbour input ports. We again observe non-classical interference signatures, with visibilities up to $\mathcal{V}_{2,4}{=} 28\pm 3\%$. The fidelities between the measured two-photon matrices and their corresponding predictions, figure~\ref{fig:corr24total}b), are $S^{d} = 97.9\pm 1\%$, and $S^{i} = 96.2\pm 0.8\%$.

\section{Discussion}
In conclusion, we have performed the first quantum walk on a general two-dimensional waveguide structure with genuine non-classical inputs. This is a significant step towards emulating Hamiltonians which can be approximated with evanescently coupled waveguides. However, we have also identified a number of obstacles which must be addressed before such an approach can be fully realised. 

First, despite the apparent good agreement between our predictions and the observed two-photon probability distributions, the two-photon visibilities---which quantify the measured non-classical effects---matched the predictions poorly. This is because the Mueller-matrix array derived from the coherent beam intensities only yields the squared absolute values of the elements of the system's unitary $U$; it does not determine the (generally complex) phase relations of the waveguide array. These phase factors could in principle be obtained by phase-sensitive coherent light tomography, as proposed in~\cite{keil2010pct,keil2011tpc}. An alternative technique~\cite{pryde2003mpb} requires a single $N$-photon input state (in this case $N=6$) and photon-number resolved detection at each output. However, generating large $N$ number states is technically difficult, and both techniques suffer from the large number of measurements required to fully characterise the six-port system. This is a perennial problem in quantum science: the exponential power granted by multi-photon walks on big lattices makes it hard to experimentally characterise system dynamics. A potential solution might be to use compressive sensing techniques, which have recently been exploited for exponentially efficient quantum process tomography \cite{shabani2009emq}. Another open question is how to deliberately control the phases in the manufacturing step of a waveguide array.

Second, while we observed significant two-photon interference visibilities, the resulting probability distributions did not exhibit a conclusively quantum signature, as quantified by the witness defined in~\cite{bromberg2009qcc}. This was most likely due to the significant polarisation-dependent coupling and loss in the circuit, leading to non-unitary evolution which failed to preserve the coherence of the input quantum state. Once we better understand the origins and mechanisms of these effects, we can exploit them to engineer devices such as polarisation-dependent couplers. It should be noted that these effects are certainly not inherent to the direct-write technique, as for example demonstrated by \cite{sansoni2010pes}. 

Future research should also focus on realising and studying effects of  decoherence in continuous-time walks. Many physical, biological or chemical systems are strongly coupled to their environment and decoherence has been suggested to be the driving factor behind phenomena such as environmentally-assisted quantum transport \cite{mohseni2008eqw}. While decoherence has been studied in discrete-time experiments \cite{broome2010dsp}, techniques for controllably introducing it to inherently robust waveguide lattices have yet to be demonstrated. 

\section*{Acknowledgments}
We thank I.~Kassal, B.~P.~Lanyon, G.~G.~Gillett, J.~C.~F.~Matthews and J.~L.~O'Brien for valuable assistance. We acknowledge financial support by ARC Discovery and Fed. Fellow programs, an IARPA-funded US Army Research Office contract and EC project QUANTIP 244026. This work was supported by the Australian Research Council through their Centres of Excellence program.

\section*{Appendix}
In the Heisenberg picture, a light field input into a waveguide in this chip is subject to the coupled-oscillator Hamiltonian~\cite{bromberg2009qcc}
\begin{equation}
H = \sum\limits_{i = 1}^{6} \beta_{i}a_{i}^{\dagger}a_{i} + \sum\limits_{i,j = 1}^{6}C_{i,j}a_{i}^{\dagger}a_{j},
\label{eq:hamiltonian}
\end{equation}
where $\beta_{i}$ is the propagation constant in waveguide $i$ and $C_{i,j}$ is the coupling constant between waveguides $i$ and $j$. The system then evolves in time according the unitary operator $U(t) = e^{-iHt/\hbar}$ and the creation operators $a_{i}^{\dagger}$ are subject to the Heisenberg equation of motion
\begin{equation}
\frac{\pr a^{\dagger}_{i}}{\pr z} = \frac{n}{c}\frac{\pr a^{\dagger}_{i}}{\pr t} =i \left[H,a^{\dagger}_{i}\right] = i\beta a^{\dagger}_{i} + i\sum\limits_{j = 1}^{6}C_{i,j}a^{\dagger}_{j},
\label{eq:evolution}
\end{equation}
which has the solution
\begin{equation}
a_{i}^{\dagger} (z) = \sum\limits_{j}(e^{izC})_{i,j}a^{\dagger}_{j}(0) = \sum\limits_{j}U_{i,j}(z)a_{j}^{\dagger}(0),
\label{eq:solution}
\end{equation}
where $C = \{ C_{i,j}\}$ is the $6\times 6$ matrix of coupling constants with diagonal entries $C_{i,i} = \beta$, and $z$ is the propagation distance along the waveguide array. Note that this evolution is equivalent to the continuous-time quantum walk formalism~\cite{farhi1998qcd}. 

The interaction length $z$ of the waveguides is chosen to match the desired run time $t$ for the emulation of the Hamiltonian. The overall response of the circuit as a 12-port beam-splitting device is then contained in the unitary matrix $U = e^{izC}$, and we can define a set of six output creation operators $b^{\dagger}_{i}$, with $b^{\dagger}_{i} = a^{\dagger}_{i}(z) = \sum\limits_{j}U_{i,j}a_{j}^{\dagger}$ for the input operators $a_{j} = a_{j}(0)$.

The coupling $C_{i,j}$ between two waveguides falls off exponentially with waveguide separation $r_{i,j}$~\cite{szameit2007cwa}, so to design an array of the type discussed in this paper, the number of waveguides, their geometry, and their separations $r_{i,j}$ are chosen to reflect the properties of the Hamiltonian under investigation. This determines the parameters $\beta_{i}$ and $C_{i,j}$. An interaction length $z$ is chosen according to the desired emulation time $t$. 

To translate the theoretical design into a real experimental system, the geometry of the circuit is analysed in a numerical electromagnetic design suite, in our case \emph{RSoft}~\cite{rsoft}. This program uses finite-difference algorithms to find solutions to Maxwell's equations in dielectrics and can be used to optimise a set of physical parameters (core diameter and refractive index contrast) which will approximate the desired evolution. This provides the link between the Hamiltonian evolution of the quantum system under study, and the physical properties of the experimental system.

We now describe the two-photon evolution in the optical system. Two temporally-indistinguishable input photons $\ket{\Psi} = a_{i}^{\dagger}a_{j}^{\dagger}\ket{0}$, for $i\neq j$, give the joint detection probability $\Gamma^{i}_{k,l}$ of finding the state $b^{\dagger}_{k}b^{\dagger}_{l}\ket{0}$ in output modes $k$ and $l$~\cite{bromberg2009qcc}:
\begin{equation}
\Gamma^{i}_{k,l} = \langle b^{\dagger}_{l}b^{\dagger}_{k}b_{k}b_{l}\rangle_{\Psi} = \frac{1}{1+\delta_{k,l}}|U_{i,k}U_{j,l} + U_{j,k}U_{i,l}|^{2}.
\label{eq:gamma} 
\end{equation}
The matrix $\Gamma^{i} = \{\Gamma^{i}_{k,l}\}$ then describes the two-photon probability distributions in all combinations of output modes. In contrast, two distinguishable photons will evolve independently and obey the statistics of Bernoulli trials. The corresponding output probability distribution $\Gamma^{d}_{k,l}$ takes the form,
\begin{equation}
\Gamma^{d}_{k,l} = \frac{1}{1+\delta_{k,l}}\left(|U_{i,k}U_{j,l}|^{2} + |U_{i,l}U_{j,k}|^{2}\right).
\label{eq:gammac}
\end{equation}
In the case of photons and electro-magnetic fields, $\Gamma^{d}_{k,l}$ represents an intensity correlation matrix $\Gamma^{d}_{k,l} = \langle I_{k}I_{l}\rangle$. The components $\Gamma^{i}_{k,l}$ and $\Gamma_{k,l}^{d}$ will differ by the factor $2\,\textrm{Re}\left[(U_{i,k}U_{j,l})^{*}(U_{i,l}U_{j,k})\right]$, which encompasses the quantum nature of indistinguishable walkers.

\begin{figure*}[t]
\setcounter{figure}{1}
\centering
\includegraphics[width=0.8\textwidth]{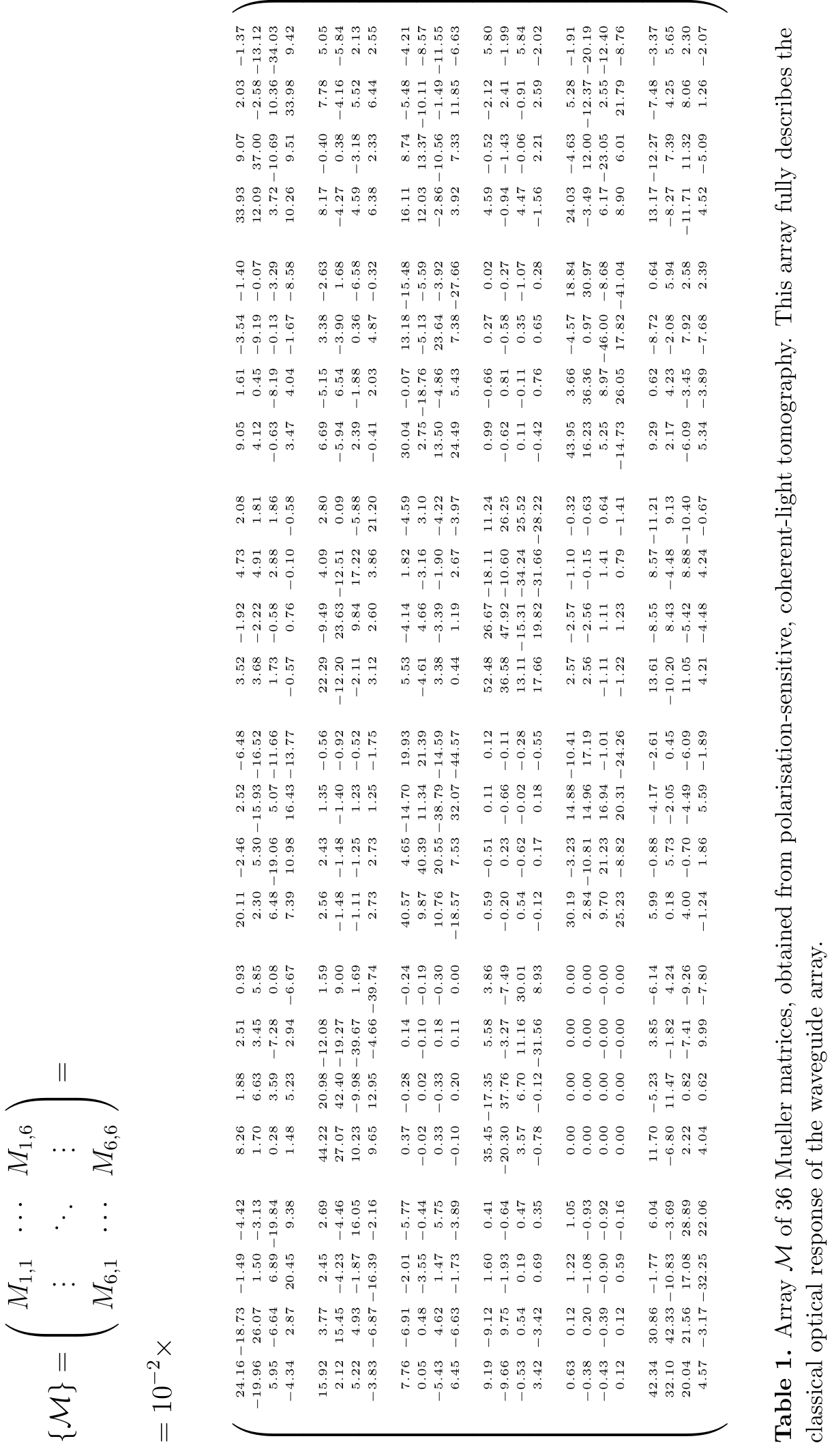}
\end{figure*}

To quantify the overlap fidelity between two probability distributions, we use $S = (\sum\limits_{k,l}\sqrt{\Gamma_{k,l}\Gamma^{p}_{k,l}})^{2}/\sum\limits_{k,l}\Gamma_{k,l}\sum\limits_{k,l}\Gamma^{p}_{k,l}$~\cite{peruzzo2010qwc}.

\subsection*{Mueller matrix}
The Mueller-matrix array $\mathcal{M}$ describes the transfer properties of the circuit in terms of the Stokes parameters $\mathcal{S}$, which describe the polarisation state of an electromagnetic field~\cite{hecht1974o}. The component $M_{i,j}$ is the $4\times 4$ Mueller matrix describing the coupling from input waveguide $j$ to output waveguide $i$. From the corresponding output Stokes parameter $\mathcal{S}_{i,j}$ for the input state $\ket{H}_{j}$, the real-valued parameter $|U_{i,j}|^{2}$ was calculated as the output transmission component in the $\ket{H}_{i}$ subspace. 

Table 1 shows the calculated Mueller-matrix array $\mathcal{M}$ of the quantum walk circuit.  The matrix not only allows us to determine the evolution matrix $U$ in the $\ket{H}$ subspace, but also quantifies the polarisation-dependent coupling and birefringence observed in the circuit, as seen in figure~\ref{fig:ellipsoids}.


\clearpage
\begin{thebibliography}{35}
\expandafter\ifx\csname natexlab\endcsname\relax\def\natexlab#1{#1}\fi
\expandafter\ifx\csname bibnamefont\endcsname\relax
  \def\bibnamefont#1{#1}\fi
\expandafter\ifx\csname bibfnamefont\endcsname\relax
  \def\bibfnamefont#1{#1}\fi
\expandafter\ifx\csname citenamefont\endcsname\relax
  \def\citenamefont#1{#1}\fi
\expandafter\ifx\csname url\endcsname\relax
  \def\url#1{\texttt{#1}}\fi
\expandafter\ifx\csname urlprefix\endcsname\relax\def\urlprefix{URL }\fi
\providecommand{\bibinfo}[2]{#2}
\providecommand{\eprint}[2][]{\url{#2}}

\bibitem[{\citenamefont{Feynman}(1982)}]{feynman1982spc}
\bibinfo{author}{\bibfnamefont{R.}~\bibnamefont{Feynman}},
  \bibinfo{journal}{International Journal of Theoretical Physics}
  \textbf{\bibinfo{volume}{21}}, \bibinfo{pages}{467} (\bibinfo{year}{1982}).

\bibitem[{\citenamefont{Lanyon et~al.}(2010)\citenamefont{Lanyon, Whitfield,
  Gillett, Goggin, de~Almeida, Kassal, Biamonte, Mohseni, Powell, Barbieri
  et~al.}}]{lanyon2010tqc}
\bibinfo{author}{\bibfnamefont{B.~P.} \bibnamefont{Lanyon}},
  \bibinfo{author}{\bibfnamefont{J.~D.} \bibnamefont{Whitfield}},
  \bibinfo{author}{\bibfnamefont{G.~G.} \bibnamefont{Gillett}},
  \bibinfo{author}{\bibfnamefont{M.~E.} \bibnamefont{Goggin}},
  \bibinfo{author}{\bibfnamefont{M.~P.} \bibnamefont{de~Almeida}},
  \bibinfo{author}{\bibfnamefont{I.}~\bibnamefont{Kassal}},
  \bibinfo{author}{\bibfnamefont{J.~D.} \bibnamefont{Biamonte}},
  \bibinfo{author}{\bibfnamefont{M.}~\bibnamefont{Mohseni}},
  \bibinfo{author}{\bibfnamefont{B.~J.} \bibnamefont{Powell}},
  \bibinfo{author}{\bibfnamefont{M.}~\bibnamefont{Barbieri}},
  \bibnamefont{et~al.}, \bibinfo{journal}{Nature Chemistry}
  \textbf{\bibinfo{volume}{2}}, \bibinfo{pages}{106} (\bibinfo{year}{2010}).

\bibitem[{\citenamefont{Du et~al.}(2010)\citenamefont{Du, Xu, Peng, Wang, Wu,
  and Lu}}]{du2010nim}
\bibinfo{author}{\bibfnamefont{J.}~\bibnamefont{Du}},
  \bibinfo{author}{\bibfnamefont{N.}~\bibnamefont{Xu}},
  \bibinfo{author}{\bibfnamefont{X.}~\bibnamefont{Peng}},
  \bibinfo{author}{\bibfnamefont{P.}~\bibnamefont{Wang}},
  \bibinfo{author}{\bibfnamefont{S.}~\bibnamefont{Wu}}, \bibnamefont{and}
  \bibinfo{author}{\bibfnamefont{D.}~\bibnamefont{Lu}}, \bibinfo{journal}{Phys.
  Rev. Lett.} \textbf{\bibinfo{volume}{104}}, \bibinfo{pages}{030502}
  (\bibinfo{year}{2010}).

\bibitem[{\citenamefont{Lloyd}(1996)}]{lloyd1996uqs}
\bibinfo{author}{\bibfnamefont{S.}~\bibnamefont{Lloyd}},
  \bibinfo{journal}{Science} \textbf{\bibinfo{volume}{273}},
  \bibinfo{pages}{1073} 
  (\bibinfo{year}{1996}).

\bibitem[{\citenamefont{Aharonov et~al.}(1993)\citenamefont{Aharonov,
  Davidovich, and Zagury}}]{aharonov1993qrw}
\bibinfo{author}{\bibfnamefont{Y.}~\bibnamefont{Aharonov}},
  \bibinfo{author}{\bibfnamefont{L.}~\bibnamefont{Davidovich}},
  \bibnamefont{and} \bibinfo{author}{\bibfnamefont{N.}~\bibnamefont{Zagury}},
  \bibinfo{journal}{Phys. Rev. A} \textbf{\bibinfo{volume}{48}},
  \bibinfo{pages}{1687} (\bibinfo{year}{1993}).

\bibitem[{\citenamefont{Kempe}(2003)}]{kempe2003qrw}
\bibinfo{author}{\bibfnamefont{J.}~\bibnamefont{Kempe}},
  \bibinfo{journal}{Contemporary Physics} \textbf{\bibinfo{volume}{44}},
  \bibinfo{pages}{307} (\bibinfo{year}{2003}).

\bibitem[{\citenamefont{M{\"u}lken and Blumen}(2011)}]{mulken2011}
\bibinfo{author}{\bibfnamefont{O.}~\bibnamefont{M{\"u}lken}} \bibnamefont{and}
  \bibinfo{author}{\bibfnamefont{A.}~\bibnamefont{Blumen}},
  \bibinfo{journal}{Arxiv preprint arXiv:1101.2572}  (\bibinfo{year}{2011}).

\bibitem[{\citenamefont{Plenio and Huelga}(2008)}]{plenio2008}
\bibinfo{author}{\bibfnamefont{M.~B.} \bibnamefont{Plenio}} \bibnamefont{and}
  \bibinfo{author}{\bibfnamefont{S.~F.} \bibnamefont{Huelga}},
  \bibinfo{journal}{New Journal of Physics} \textbf{\bibinfo{volume}{10}},
  \bibinfo{pages}{113019} (\bibinfo{year}{2008}).

\bibitem[{\citenamefont{Mohseni et~al.}(2008)\citenamefont{Mohseni, Rebentrost,
  Lloyd, and Aspuru-Guzik}}]{mohseni2008eqw}
\bibinfo{author}{\bibfnamefont{M.}~\bibnamefont{Mohseni}},
  \bibinfo{author}{\bibfnamefont{P.}~\bibnamefont{Rebentrost}},
  \bibinfo{author}{\bibfnamefont{S.}~\bibnamefont{Lloyd}}, \bibnamefont{and}
  \bibinfo{author}{\bibfnamefont{A.}~\bibnamefont{Aspuru-Guzik}},
  \bibinfo{journal}{J. Chem. Phys.} \textbf{\bibinfo{volume}{129}},
  \bibinfo{pages}{174106} (\bibinfo{year}{2008}).

\bibitem[{\citenamefont{Karski et~al.}(2009)\citenamefont{Karski, Forster,
  Choi, Steffen, Alt, Meschede, and Widera}}]{karski2009qwp}
\bibinfo{author}{\bibfnamefont{M.}~\bibnamefont{Karski}},
  \bibinfo{author}{\bibfnamefont{L.}~\bibnamefont{Forster}},
  \bibinfo{author}{\bibfnamefont{J.-M.} \bibnamefont{Choi}},
  \bibinfo{author}{\bibfnamefont{A.}~\bibnamefont{Steffen}},
  \bibinfo{author}{\bibfnamefont{W.}~\bibnamefont{Alt}},
  \bibinfo{author}{\bibfnamefont{D.}~\bibnamefont{Meschede}}, \bibnamefont{and}
  \bibinfo{author}{\bibfnamefont{A.}~\bibnamefont{Widera}},
  \bibinfo{journal}{Science} \textbf{\bibinfo{volume}{325}},
  \bibinfo{pages}{174} (\bibinfo{year}{2009}).

\bibitem[{\citenamefont{Z{\"a}hringer et~al.}(2010)\citenamefont{Z{\"a}hringer,
  Kirchmair, Gerritsma, Solano, Blatt, and Roos}}]{zaehringer2010rqw}
\bibinfo{author}{\bibfnamefont{F.}~\bibnamefont{Z{\"a}hringer}},
  \bibinfo{author}{\bibfnamefont{G.}~\bibnamefont{Kirchmair}},
  \bibinfo{author}{\bibfnamefont{R.}~\bibnamefont{Gerritsma}},
  \bibinfo{author}{\bibfnamefont{E.}~\bibnamefont{Solano}},
  \bibinfo{author}{\bibfnamefont{R.}~\bibnamefont{Blatt}}, \bibnamefont{and}
  \bibinfo{author}{\bibfnamefont{C.~F.} \bibnamefont{Roos}},
  \bibinfo{journal}{Phys. Rev. Lett.} \textbf{\bibinfo{volume}{104}},
  \bibinfo{pages}{100503} (\bibinfo{year}{2010}).

\bibitem[{\citenamefont{Schmitz et~al.}(2009)\citenamefont{Schmitz, Matjeschk,
  Schneider, Glueckert, Enderlein, Huber, and Schaetz}}]{schmitz2009qwt}
\bibinfo{author}{\bibfnamefont{H.}~\bibnamefont{Schmitz}},
  \bibinfo{author}{\bibfnamefont{R.}~\bibnamefont{Matjeschk}},
  \bibinfo{author}{\bibfnamefont{C.}~\bibnamefont{Schneider}},
  \bibinfo{author}{\bibfnamefont{J.}~\bibnamefont{Glueckert}},
  \bibinfo{author}{\bibfnamefont{M.}~\bibnamefont{Enderlein}},
  \bibinfo{author}{\bibfnamefont{T.}~\bibnamefont{Huber}}, \bibnamefont{and}
  \bibinfo{author}{\bibfnamefont{T.}~\bibnamefont{Schaetz}},
  \bibinfo{journal}{Phys. Rev. Lett.} \textbf{\bibinfo{volume}{103}},
  \bibinfo{pages}{090504} (\bibinfo{year}{2009}).

\bibitem[{\citenamefont{Schreiber et~al.}(2010)\citenamefont{Schreiber,
  Cassemiro, Poto{\v{c}}ek, G{\'a}bris, Mosley, Andersson, Jex, and
  Silberhorn}}]{schreiber2010pwl}
\bibinfo{author}{\bibfnamefont{A.}~\bibnamefont{Schreiber}},
  \bibinfo{author}{\bibfnamefont{K.~N.} \bibnamefont{Cassemiro}},
  \bibinfo{author}{\bibfnamefont{V.}~\bibnamefont{Poto{\v{c}}ek}},
  \bibinfo{author}{\bibfnamefont{A.}~\bibnamefont{G{\'a}bris}},
  \bibinfo{author}{\bibfnamefont{P.~J.} \bibnamefont{Mosley}},
  \bibinfo{author}{\bibfnamefont{E.}~\bibnamefont{Andersson}},
  \bibinfo{author}{\bibfnamefont{I.}~\bibnamefont{Jex}}, \bibnamefont{and}
  \bibinfo{author}{\bibfnamefont{C.}~\bibnamefont{Silberhorn}},
  \bibinfo{journal}{Phys. Rev. Lett.} \textbf{\bibinfo{volume}{104}},
  \bibinfo{pages}{50502} (\bibinfo{year}{2010}), ISSN
  \bibinfo{issn}{1079-7114}.

\bibitem[{\citenamefont{Broome et~al.}(2010)\citenamefont{Broome, Fedrizzi,
  Lanyon, Kassal, Aspuru-Guzik, and White}}]{broome2010dsp}
\bibinfo{author}{\bibfnamefont{M.~A.} \bibnamefont{Broome}},
  \bibinfo{author}{\bibfnamefont{A.}~\bibnamefont{Fedrizzi}},
  \bibinfo{author}{\bibfnamefont{B.~P.} \bibnamefont{Lanyon}},
  \bibinfo{author}{\bibfnamefont{I.}~\bibnamefont{Kassal}},
  \bibinfo{author}{\bibfnamefont{A.}~\bibnamefont{Aspuru-Guzik}},
  \bibnamefont{and} \bibinfo{author}{\bibfnamefont{A.~G.} \bibnamefont{White}},
  \bibinfo{journal}{Phys. Rev. Lett.} \textbf{\bibinfo{volume}{104}},
  \bibinfo{pages}{153602} (\bibinfo{year}{2010}), ISSN
  \bibinfo{issn}{1079-7114}.

\bibitem[{\citenamefont{Farhi and Gutmann}(1998)}]{farhi1998qcd}
\bibinfo{author}{\bibfnamefont{E.}~\bibnamefont{Farhi}} \bibnamefont{and}
  \bibinfo{author}{\bibfnamefont{S.}~\bibnamefont{Gutmann}},
  \bibinfo{journal}{Physical Review A} \textbf{\bibinfo{volume}{58}},
  \bibinfo{pages}{915} (\bibinfo{year}{1998}), ISSN \bibinfo{issn}{1094-1622}.

\bibitem[{\citenamefont{Bromberg et~al.}(2009)\citenamefont{Bromberg, Lahini,
  Morandotti, and Silberberg}}]{bromberg2009qcc}
\bibinfo{author}{\bibfnamefont{Y.}~\bibnamefont{Bromberg}},
  \bibinfo{author}{\bibfnamefont{Y.}~\bibnamefont{Lahini}},
  \bibinfo{author}{\bibfnamefont{R.}~\bibnamefont{Morandotti}},
  \bibnamefont{and}
  \bibinfo{author}{\bibfnamefont{Y.}~\bibnamefont{Silberberg}},
  \bibinfo{journal}{Phys. Rev. Lett.} \textbf{\bibinfo{volume}{102}},
  \bibinfo{pages}{253904} (\bibinfo{year}{2009}).

\bibitem[{\citenamefont{Marshall et~al.}(2009)\citenamefont{Marshall, Politi,
  Matthews, Dekker, Ams, Withford, and O'Brien}}]{marshall2009lww}
\bibinfo{author}{\bibfnamefont{G.~D.} \bibnamefont{Marshall}},
  \bibinfo{author}{\bibfnamefont{A.}~\bibnamefont{Politi}},
  \bibinfo{author}{\bibfnamefont{J.~C.~F.} \bibnamefont{Matthews}},
  \bibinfo{author}{\bibfnamefont{P.}~\bibnamefont{Dekker}},
  \bibinfo{author}{\bibfnamefont{M.}~\bibnamefont{Ams}},
  \bibinfo{author}{\bibfnamefont{M.~J.} \bibnamefont{Withford}},
  \bibnamefont{and} \bibinfo{author}{\bibfnamefont{J.~L.}
  \bibnamefont{O'Brien}}, \bibinfo{journal}{Opt. Express}
  \textbf{\bibinfo{volume}{17}}, \bibinfo{pages}{12546} (\bibinfo{year}{2009}).

\bibitem[{\citenamefont{Hein and Tanner}(2009)}]{hein2009wca}
\bibinfo{author}{\bibfnamefont{B.}~\bibnamefont{Hein}} \bibnamefont{and}
  \bibinfo{author}{\bibfnamefont{G.}~\bibnamefont{Tanner}},
  \bibinfo{journal}{Physical review letters} \textbf{\bibinfo{volume}{103}},
  \bibinfo{pages}{260501} (\bibinfo{year}{2009}), ISSN
  \bibinfo{issn}{1079-7114}.

\bibitem[{\citenamefont{Bulger et~al.}(2008)\citenamefont{Bulger, Freckleton,
  and Twamley}}]{bulger2008pdc}
\bibinfo{author}{\bibfnamefont{D.}~\bibnamefont{Bulger}},
  \bibinfo{author}{\bibfnamefont{J.}~\bibnamefont{Freckleton}},
  \bibnamefont{and} \bibinfo{author}{\bibfnamefont{J.}~\bibnamefont{Twamley}},
  \bibinfo{journal}{New Journal of Physics} \textbf{\bibinfo{volume}{10}},
  \bibinfo{pages}{093014} (\bibinfo{year}{2008}).

\bibitem[{\citenamefont{Kitagawa et~al.}(2010)\citenamefont{Kitagawa, Rudner,
  Berg, and Demler}}]{kitagawa2010etp}
\bibinfo{author}{\bibfnamefont{T.}~\bibnamefont{Kitagawa}},
  \bibinfo{author}{\bibfnamefont{M.}~\bibnamefont{Rudner}},
  \bibinfo{author}{\bibfnamefont{E.}~\bibnamefont{Berg}}, \bibnamefont{and}
  \bibinfo{author}{\bibfnamefont{E.}~\bibnamefont{Demler}},
  \bibinfo{journal}{Physical Review A} \textbf{\bibinfo{volume}{82}},
  \bibinfo{pages}{33429} (\bibinfo{year}{2010}), ISSN
  \bibinfo{issn}{0556-2791}.

\bibitem[{\citenamefont{Szameit et~al.}(2007)\citenamefont{Szameit, Dreisow,
  Pertsch, Nolte, and T{\"u}nnermann}}]{szameit2007cwa}
\bibinfo{author}{\bibfnamefont{A.}~\bibnamefont{Szameit}},
  \bibinfo{author}{\bibfnamefont{F.}~\bibnamefont{Dreisow}},
  \bibinfo{author}{\bibfnamefont{T.}~\bibnamefont{Pertsch}},
  \bibinfo{author}{\bibfnamefont{S.}~\bibnamefont{Nolte}}, \bibnamefont{and}
  \bibinfo{author}{\bibfnamefont{A.}~\bibnamefont{T{\"u}nnermann}},
  \bibinfo{journal}{Opt. Express} \textbf{\bibinfo{volume}{15}},
  \bibinfo{pages}{1579} (\bibinfo{year}{2007}).

\bibitem[{\citenamefont{R{\"o}pke et~al.}(2007)\citenamefont{R{\"o}pke,
  Bartelt, Unger, Schuster, and Kobelke}}]{roepke2007tdh}
\bibinfo{author}{\bibfnamefont{U.}~\bibnamefont{R{\"o}pke}},
  \bibinfo{author}{\bibfnamefont{H.}~\bibnamefont{Bartelt}},
  \bibinfo{author}{\bibfnamefont{S.}~\bibnamefont{Unger}},
  \bibinfo{author}{\bibfnamefont{K.}~\bibnamefont{Schuster}}, \bibnamefont{and}
  \bibinfo{author}{\bibfnamefont{J.}~\bibnamefont{Kobelke}},
  \bibinfo{journal}{Optics Express} \textbf{\bibinfo{volume}{15}},
  \bibinfo{pages}{6894} (\bibinfo{year}{2007}), ISSN \bibinfo{issn}{1094-4087}.

\bibitem[{\citenamefont{Keil et~al.}(2010)\citenamefont{Keil, Szameit, Dreisow,
  Heinrich, Nolte, and T{\"u}nnermann}}]{keil2010pct}
\bibinfo{author}{\bibfnamefont{R.}~\bibnamefont{Keil}},
  \bibinfo{author}{\bibfnamefont{A.}~\bibnamefont{Szameit}},
  \bibinfo{author}{\bibfnamefont{F.}~\bibnamefont{Dreisow}},
  \bibinfo{author}{\bibfnamefont{M.}~\bibnamefont{Heinrich}},
  \bibinfo{author}{\bibfnamefont{S.}~\bibnamefont{Nolte}}, \bibnamefont{and}
  \bibinfo{author}{\bibfnamefont{A.}~\bibnamefont{T{\"u}nnermann}},
  \bibinfo{journal}{Physical Review A} \textbf{\bibinfo{volume}{81}},
  \bibinfo{pages}{23834} (\bibinfo{year}{2010}), ISSN
  \bibinfo{issn}{1094-1622}.

\bibitem[{\citenamefont{Rohde et~al.}(2011)\citenamefont{Rohde, Schreiber,
  Stefanak, Jex, and Silberhorn}}]{rohde2011mdt}
\bibinfo{author}{\bibfnamefont{P.~P.} \bibnamefont{Rohde}},
  \bibinfo{author}{\bibfnamefont{A.}~\bibnamefont{Schreiber}},
  \bibinfo{author}{\bibfnamefont{M.}~\bibnamefont{Stefanak}},
  \bibinfo{author}{\bibfnamefont{I.}~\bibnamefont{Jex}}, \bibnamefont{and}
  \bibinfo{author}{\bibfnamefont{C.}~\bibnamefont{Silberhorn}},
  \bibinfo{journal}{New Journal of Physics} \textbf{\bibinfo{volume}{13}},
  \bibinfo{pages}{013001} (\bibinfo{year}{2011}).

\bibitem[{\citenamefont{Peruzzo et~al.}(2010)\citenamefont{Peruzzo, Lobino,
  Matthews, Matsuda, Politi, Poulios, Zhou, Lahini, Ismail, W{\"o}rhoff
  et~al.}}]{peruzzo2010qwc}
\bibinfo{author}{\bibfnamefont{A.}~\bibnamefont{Peruzzo}},
  \bibinfo{author}{\bibfnamefont{M.}~\bibnamefont{Lobino}},
  \bibinfo{author}{\bibfnamefont{J.}~\bibnamefont{Matthews}},
  \bibinfo{author}{\bibfnamefont{N.}~\bibnamefont{Matsuda}},
  \bibinfo{author}{\bibfnamefont{A.}~\bibnamefont{Politi}},
  \bibinfo{author}{\bibfnamefont{K.}~\bibnamefont{Poulios}},
  \bibinfo{author}{\bibfnamefont{X.}~\bibnamefont{Zhou}},
  \bibinfo{author}{\bibfnamefont{Y.}~\bibnamefont{Lahini}},
  \bibinfo{author}{\bibfnamefont{N.}~\bibnamefont{Ismail}},
  \bibinfo{author}{\bibfnamefont{K.}~\bibnamefont{W{\"o}rhoff}},
  \bibnamefont{et~al.}, \bibinfo{journal}{Science}
  \textbf{\bibinfo{volume}{329}}, \bibinfo{pages}{1500} (\bibinfo{year}{2010}).

\bibitem[{\citenamefont{Davis et~al.}(1996)\citenamefont{Davis, Miura,
  Sugimoto, and Hirao}}]{davis1996}
\bibinfo{author}{\bibfnamefont{K.~M.} \bibnamefont{Davis}},
  \bibinfo{author}{\bibfnamefont{K.}~\bibnamefont{Miura}},
  \bibinfo{author}{\bibfnamefont{N.}~\bibnamefont{Sugimoto}}, \bibnamefont{and}
  \bibinfo{author}{\bibfnamefont{K.}~\bibnamefont{Hirao}},
  \bibinfo{journal}{Optics Letters} \textbf{\bibinfo{volume}{21}},
  \bibinfo{pages}{1729} (\bibinfo{year}{1996}).

\bibitem[{\citenamefont{Nolte et~al.}(2003)\citenamefont{Nolte, Will, Burghoff,
  and a.~T{\"u}nnermann}}]{nolte2003}
\bibinfo{author}{\bibfnamefont{S.}~\bibnamefont{Nolte}},
  \bibinfo{author}{\bibfnamefont{M.}~\bibnamefont{Will}},
  \bibinfo{author}{\bibfnamefont{J.}~\bibnamefont{Burghoff}}, \bibnamefont{and}
  \bibinfo{author}{\bibnamefont{a.~T{\"u}nnermann}}, \bibinfo{journal}{Applied
  Physics A} \textbf{\bibinfo{volume}{77}}, \bibinfo{pages}{109}
  (\bibinfo{year}{2003}).

\bibitem[{\citenamefont{Minford et~al.}(1982)\citenamefont{Minford, Korotky,
  and Alferness}}]{minford1982}
\bibinfo{author}{\bibfnamefont{W.~J.} \bibnamefont{Minford}},
  \bibinfo{author}{\bibfnamefont{S.~K.} \bibnamefont{Korotky}},
  \bibnamefont{and} \bibinfo{author}{\bibfnamefont{R.~C.}
  \bibnamefont{Alferness}}, \bibinfo{journal}{IEEE Journal of Quantum
  Electronics} \textbf{\bibinfo{volume}{QE-18}}, \bibinfo{pages}{1802}
  (\bibinfo{year}{1982}).

\bibitem[{\citenamefont{Hecht and Zajac}(1974)}]{hecht1974o}
\bibinfo{author}{\bibfnamefont{E.}~\bibnamefont{Hecht}} \bibnamefont{and}
  \bibinfo{author}{\bibfnamefont{A.}~\bibnamefont{Zajac}},
  \bibinfo{journal}{Addison-Wesley series in physics}  (\bibinfo{year}{1974}).

\bibitem[{\citenamefont{Hong et~al.}(1987)\citenamefont{Hong, Ou, and
  Mandel}}]{hong1987hom}
\bibinfo{author}{\bibfnamefont{C.~K.} \bibnamefont{Hong}},
  \bibinfo{author}{\bibfnamefont{Z.~Y.} \bibnamefont{Ou}}, \bibnamefont{and}
  \bibinfo{author}{\bibfnamefont{L.}~\bibnamefont{Mandel}},
  \bibinfo{journal}{Phys. Rev. Lett.} \textbf{\bibinfo{volume}{59}},
  \bibinfo{pages}{2044} (\bibinfo{year}{1987}).

\bibitem[{\citenamefont{Keil et~al.}(2011)\citenamefont{Keil, Dreisow,
  Heinrich, T{\"u}nnermann, Nolte, and Szameit}}]{keil2011tpc}
\bibinfo{author}{\bibfnamefont{R.}~\bibnamefont{Keil}},
  \bibinfo{author}{\bibfnamefont{F.}~\bibnamefont{Dreisow}},
  \bibinfo{author}{\bibfnamefont{M.}~\bibnamefont{Heinrich}},
  \bibinfo{author}{\bibfnamefont{A.}~\bibnamefont{T{\"u}nnermann}},
  \bibinfo{author}{\bibfnamefont{S.}~\bibnamefont{Nolte}}, \bibnamefont{and}
  \bibinfo{author}{\bibfnamefont{A.}~\bibnamefont{Szameit}},
  \bibinfo{journal}{Phys. Rev. A} \textbf{\bibinfo{volume}{83}},
  \bibinfo{pages}{013808} (\bibinfo{year}{2011}).

\bibitem[{\citenamefont{Pryde and White}(2003)}]{pryde2003mpb}
\bibinfo{author}{\bibfnamefont{G.~J.} \bibnamefont{Pryde}} \bibnamefont{and}
  \bibinfo{author}{\bibfnamefont{A.~G.} \bibnamefont{White}},
  \bibinfo{journal}{Phys. Rev. A} \textbf{\bibinfo{volume}{68}},
  \bibinfo{pages}{052315} (\bibinfo{year}{2003}).

\bibitem[{\citenamefont{Shabani et~al.}(2009)\citenamefont{Shabani, Kosut,
  Mohseni, Rabitz, Broome, Almeida, Fedrizzi, and White}}]{shabani2009emq}
\bibinfo{author}{\bibfnamefont{A.}~\bibnamefont{Shabani}},
  \bibinfo{author}{\bibfnamefont{R.}~\bibnamefont{Kosut}},
  \bibinfo{author}{\bibfnamefont{M.}~\bibnamefont{Mohseni}},
  \bibinfo{author}{\bibfnamefont{H.}~\bibnamefont{Rabitz}},
  \bibinfo{author}{\bibfnamefont{M.}~\bibnamefont{Broome}},
  \bibinfo{author}{\bibfnamefont{M.}~\bibnamefont{Almeida}},
  \bibinfo{author}{\bibfnamefont{A.}~\bibnamefont{Fedrizzi}}, \bibnamefont{and}
  \bibinfo{author}{\bibfnamefont{A.}~\bibnamefont{White}},
  \bibinfo{journal}{Arxiv preprint arXiv:0910.5498}  (\bibinfo{year}{2009}).

\bibitem[{\citenamefont{Sansoni et~al.}(2010)\citenamefont{Sansoni, Sciarrino,
  Vallone, Mataloni, Crespi, Ramponi, and Osellame}}]{sansoni2010pes}
\bibinfo{author}{\bibfnamefont{L.}~\bibnamefont{Sansoni}},
  \bibinfo{author}{\bibfnamefont{F.}~\bibnamefont{Sciarrino}},
  \bibinfo{author}{\bibfnamefont{G.}~\bibnamefont{Vallone}},
  \bibinfo{author}{\bibfnamefont{P.}~\bibnamefont{Mataloni}},
  \bibinfo{author}{\bibfnamefont{A.}~\bibnamefont{Crespi}},
  \bibinfo{author}{\bibfnamefont{R.}~\bibnamefont{Ramponi}}, \bibnamefont{and}
  \bibinfo{author}{\bibfnamefont{R.}~\bibnamefont{Osellame}},
  \bibinfo{journal}{Phys. Rev. Lett.} \textbf{\bibinfo{volume}{105}},
  \bibinfo{pages}{200503} (\bibinfo{year}{2010}).

\bibitem{rsoft}
 {\em http://www.rsoftdesign.com/\/}

\end{thebibliography}
\end{document}